\documentclass[a4paper,
               keeplastbox,   
               ]{jacow}
\usepackage{pdfpages,multirow,ragged2e} %
\makeatletter%
	\ifboolexpr{bool{xetex}}
	 {\renewcommand{\Gin@extensions}{.pdf,%
	                    .png,.jpg,.bmp,.pict,.tif,.psd,.mac,.sga,.tga,.gif,%
	                    .eps,.ps,%
	                    }}{}
\makeatother

\ifboolexpr{bool{xetex} or bool{luatex}} 
 {}                                      
 {\usepackage[utf8]{inputenc}}           

\usepackage[USenglish]{babel}
\usepackage{doi}

\ifboolexpr{bool{jacowbiblatex}}%
 {%
  \addbibresource{jacow-test.bib}
  \addbibresource{biblatex-examples.bib}
 }{}
\begin{document}



\title{LLRF for PolFEL Accelerator\thanks{Work supported by National Information Processing Institute, \texttt{https://opi.org.pl/}}}

\author{J. Szewiński\thanks{J.Szewinski@ncbj.gov.pl}, P. Bartoszek, K. Chmielewski,
  T. Kowalski, P. Krawczyk, R. Nietubyć,\\
  D. Rybka, M. Sitek, National Centre for Nuclear Research, Otwock-Świerk, Poland}
	
\maketitle

\begin{abstract}
  PolFEL stands for Polish Free Electron Laser, the first FEL research infrastructure in Poland.
  This facility is under development, and it will operate in three wavelength ranges:
  IR, THz and VUV, using different types of undulators.
  Machine will be driven by 200 MeV linear superconducting accelerator,
  which will operate in both, pulsed wave (PW) and continous wave (CW) modes.
  This paper will describe the concept, current status and the first results of the LLRF systems development. 
\end{abstract}

\section{LOW-LEVEL RF SYSTEM OVERVIEW}
High speed and high bandwidth ADCs makes possible to sample directly the RF signal of the
frequency 1.3 GHz. Well known and also evaluated \cite{geng,okada} for this purpose is Texas Instruments ADS5474,
which input bandwidth covers range up to 1.4 GHz. Possibility of direct RF sampling allows
to significantly simplify the LLRF hardware.
The key part of the direct sampling LLRF system will be the clocking solution for the ADCs, but since the whole clock tree will be integrated into the ADC board, it can be optimized for this particular application.

The components of the PolFEL LLRF system are similar to the ones used at X-FEL\cite{xfel_llrf} because of the same fundamental frequency \SI{1.3}{GHz}, but the layout of the system is more like the one used at ESS\cite{ess_llrf},
because ESS operates also in single cavity regulation mode.

Configuration used at ESS for controlling single cavity occupies 3 slots in the MTCA chassis,
and results in total number of 6 devices (3xAMC + 3xRTM) for single cavity.
One slot is occupied by the main LLRF Controller, which uses both boards:
AMC with FPGA and data converters, and RTM with the downconverters and vector modulator.
Other two slots are occupied by the piezo controller, and LO clock signal generator.

\begin{figure}[htb!]
  \centering
  \includegraphics[width=.4\textwidth]{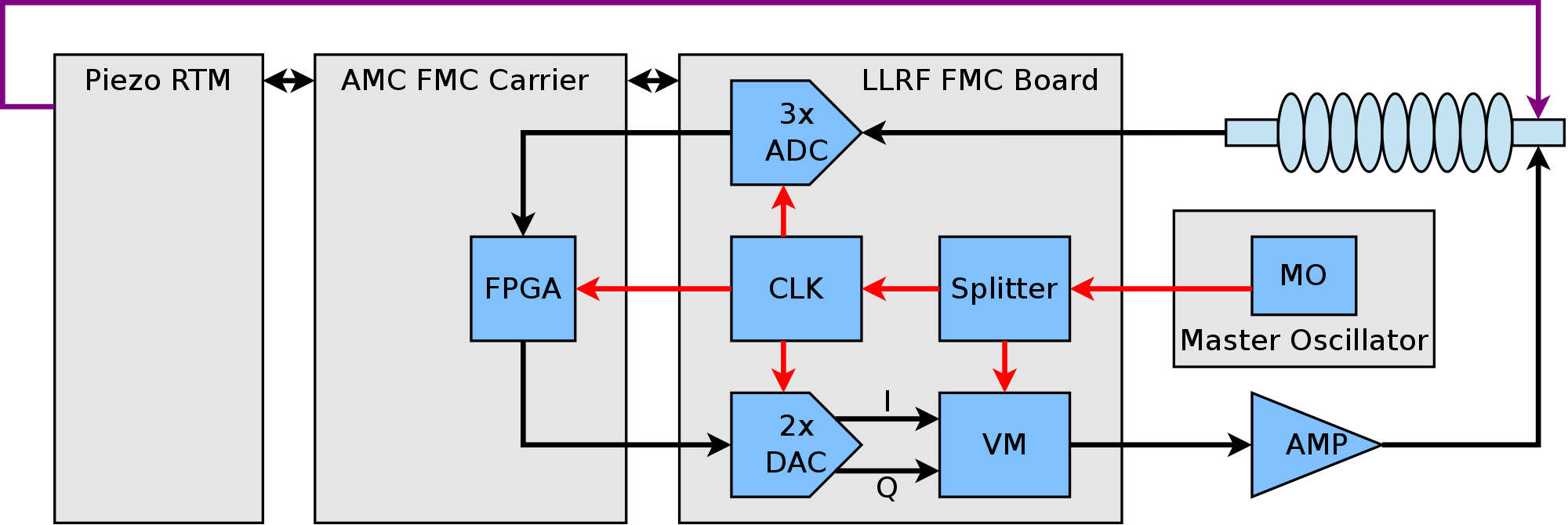}
  \caption{PolFEL LLRF System architecture}
  \label{fig:polfel_llrf}
\end{figure}

The concept of the PolFEL LLRF controller (Fig. \ref{fig:polfel_llrf}) is much simpler,
for single cavity control single MTCA chassis slot is occupied.
From the rear side of the slot the Piezo RTM will be placed, and from the front side an AMC 
FMC Carrier will be used. All LLRF specific infrastructure will be placed on the custom
dual FMC board. With respect to the amount of connected I/O pins in the FMC connectors,
any MTCA.4 FMC carrier can be used. This configuration does not require down-converters,
so separate LO generation device is not required as well.

The function of the LO generation is performed by proper circuitry integrated in the FMC board along with the ADCs and DACs. The ADC sampling clock is generated directly from the 1.3 RF signal, and the distance from
RF input to the ADC or vector modulator is less than 10 cm. All clock distribution for a single LLRF system will be
made on the single PCB.

\subsection{INITIAL TESTS OF LLRF WITH THE COPPER CAVITY}
To evaluate described concept, a test setup has been assembled.
To make tests as much similar to the final applucation, a 1.3 GHz, 3-cell, copper
cavity has been used.

\begin{figure}
  \centering
  \includegraphics[width=0.4\textwidth]{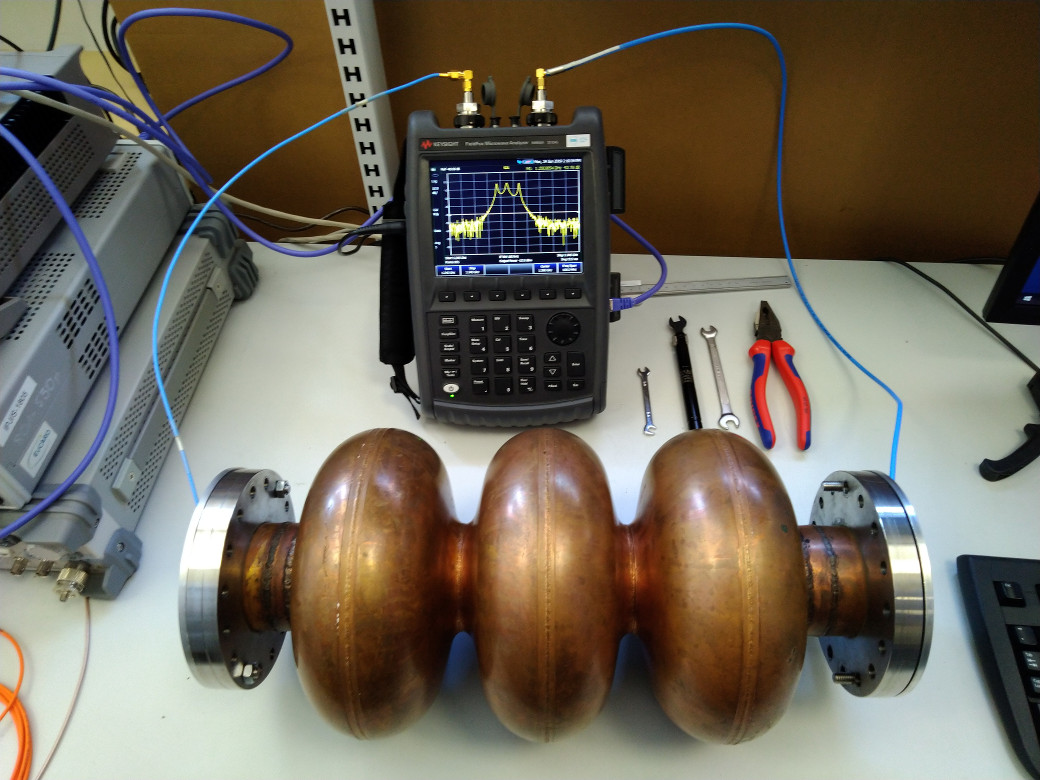}
  \caption{Cavity characterized using VNA}
  \label{fig:cavity_vna}
\end{figure}

  \begin{figure}
  \includegraphics[width=0.48\textwidth]{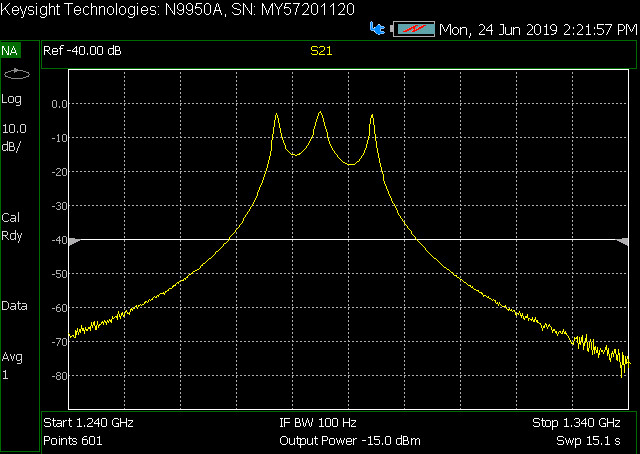}\\
  \caption{Cavity amplitude response}
  \label{fig:cav_response}
  \end{figure}


As a first step, the cavity has been measured and couplers tuned using Vector-Network Analyzer (Figs. \ref{fig:cavity_vna},\ref{fig:cav_response}).
In the next step, the field in the cavity has been excited using the RF generator on one side,
while the signal from the other coupler was connected to FPGA based system (Xilinx KC705 evaluation board) with ADS5474 ADC attached on the Curtiss-Wright ADC511 FMC mezzanine.
Using digital I/Q detection, it was possible in the FPGA to restore the amplitude and phase of the cavity field.

In order to close the feedback loop, vector modulator was needed. For the purpose the FMC board with dual-channel DAC and vector-modulator has been designed, and manufactured.

\begin{figure}[htb!]
  \centering
  \includegraphics[width=.4\textwidth]{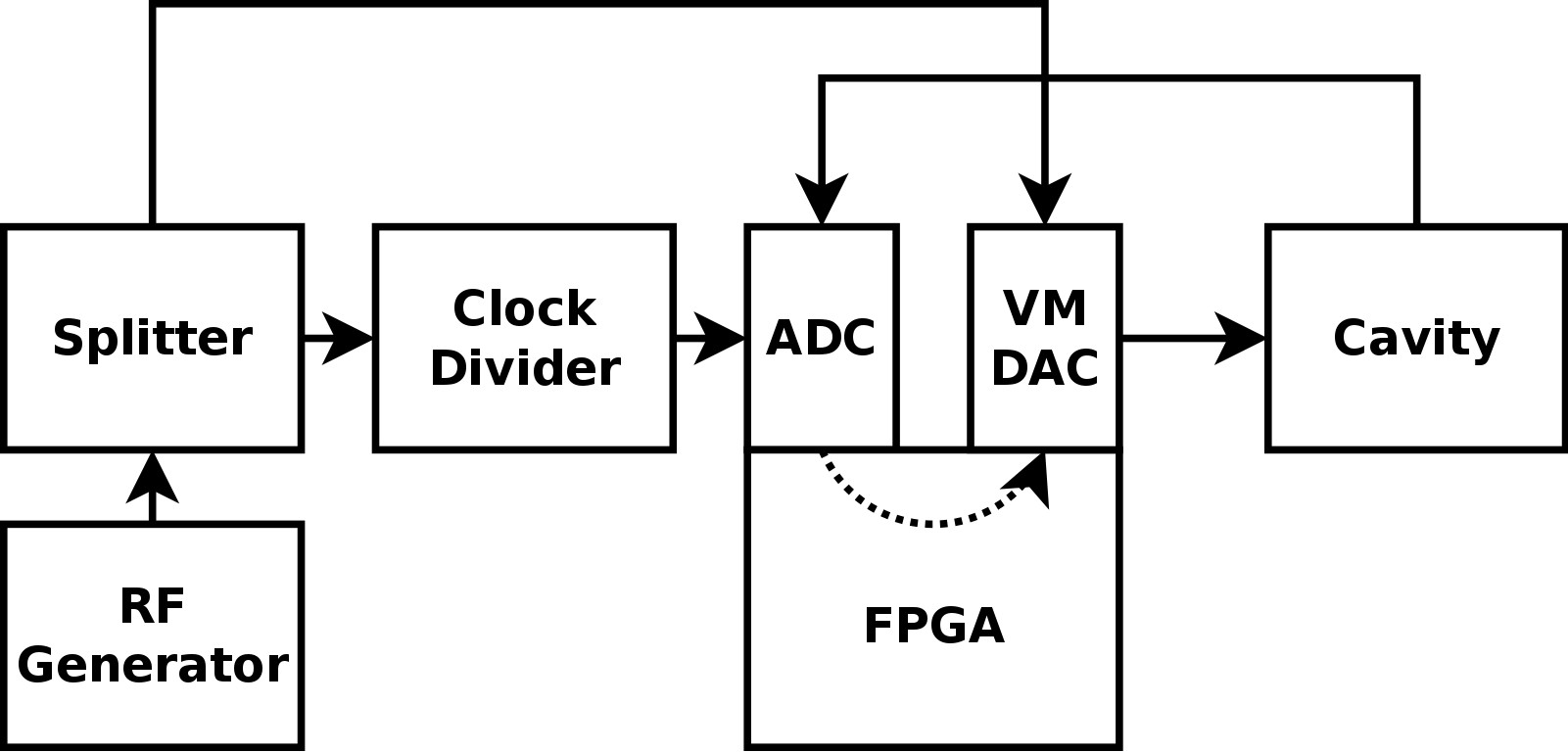}\\
  \caption{The scheme of the test setup}
  \label{fig:scheme}
\end{figure}

Having all components available, the complete setup has been assembled (Fig.~\ref{fig:scheme}).
Signal from the RF generator has been split and delivered to vector modulator as source RF signal to be modulated,
and to clock synthesizer, which generates frequencies suitable for ADCs and DACs. Clock synthesizer can additionally provide clock signal
to the FPGA device using FMC\_M2C signals, but this was not necessary because ADCs and DACs provides clock synchronized with their data.

\begin{figure}[htb!]
\includegraphics[width=0.48\textwidth]{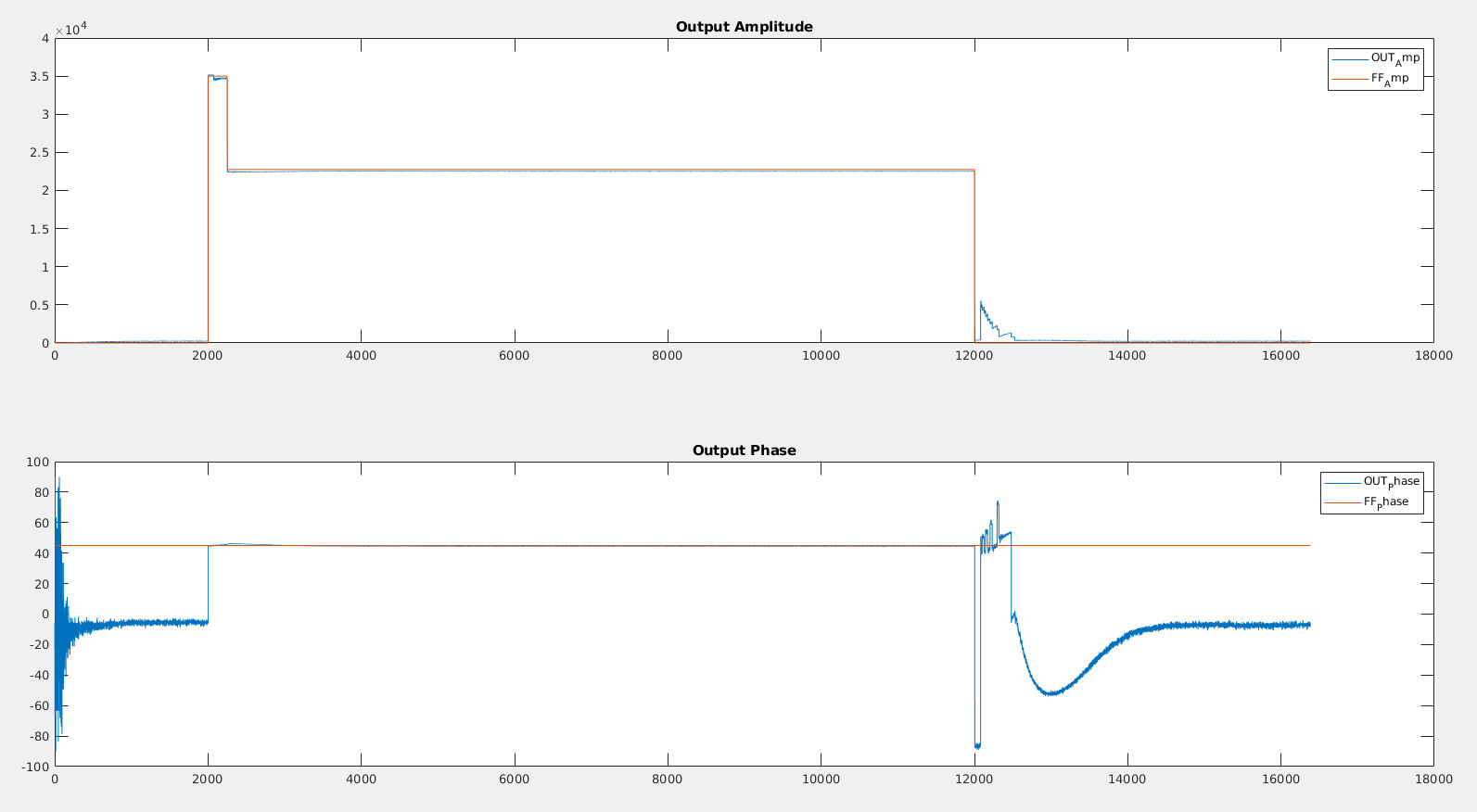}
\caption{Controller output}
\label{fig:output}
\end{figure}

\begin{figure}[htb!]
\includegraphics[width=0.48\textwidth]{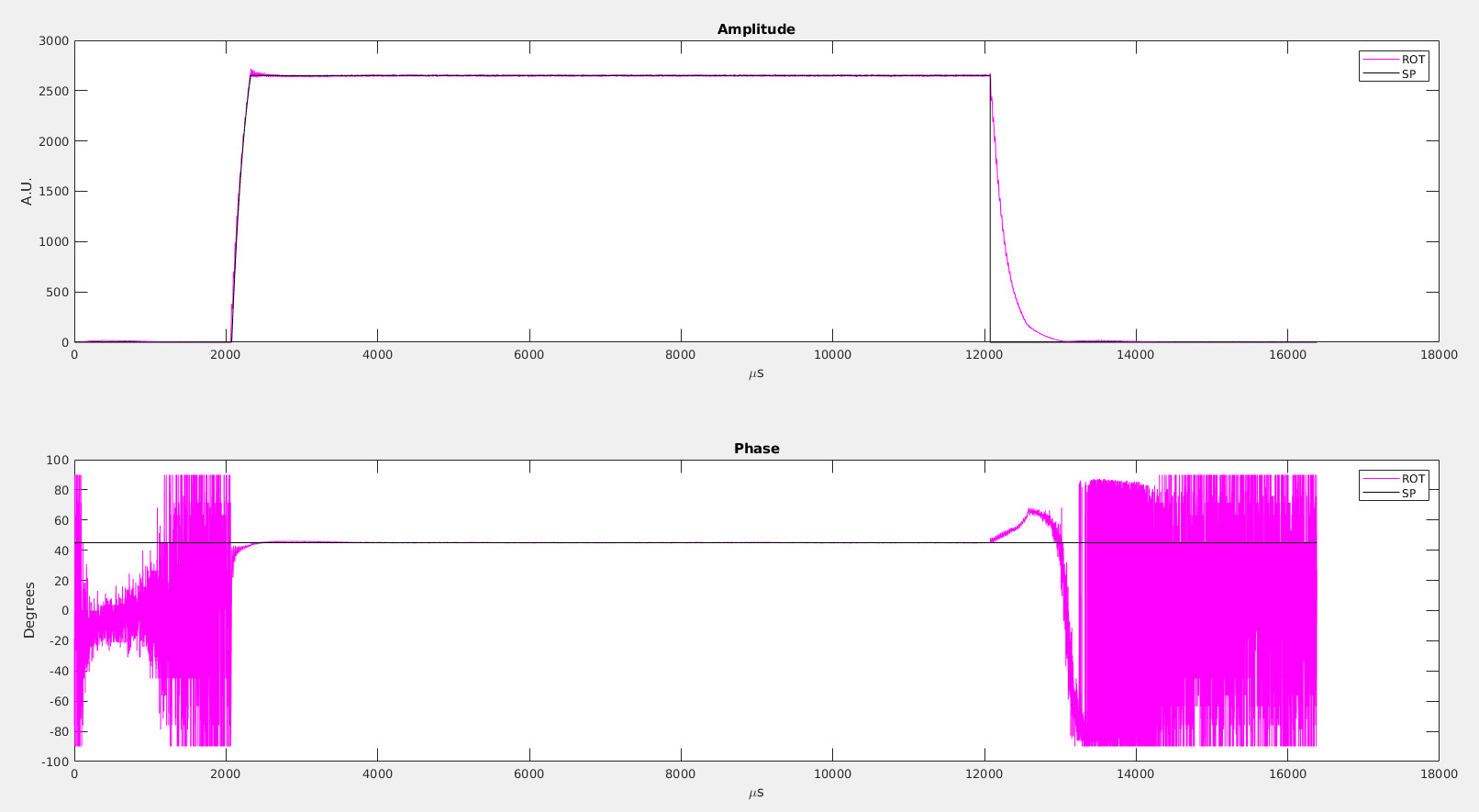}
\caption{Controller input}
\label{fig:input}
\end{figure}

\begin{figure}[htb!]
\includegraphics[width=0.48\textwidth]{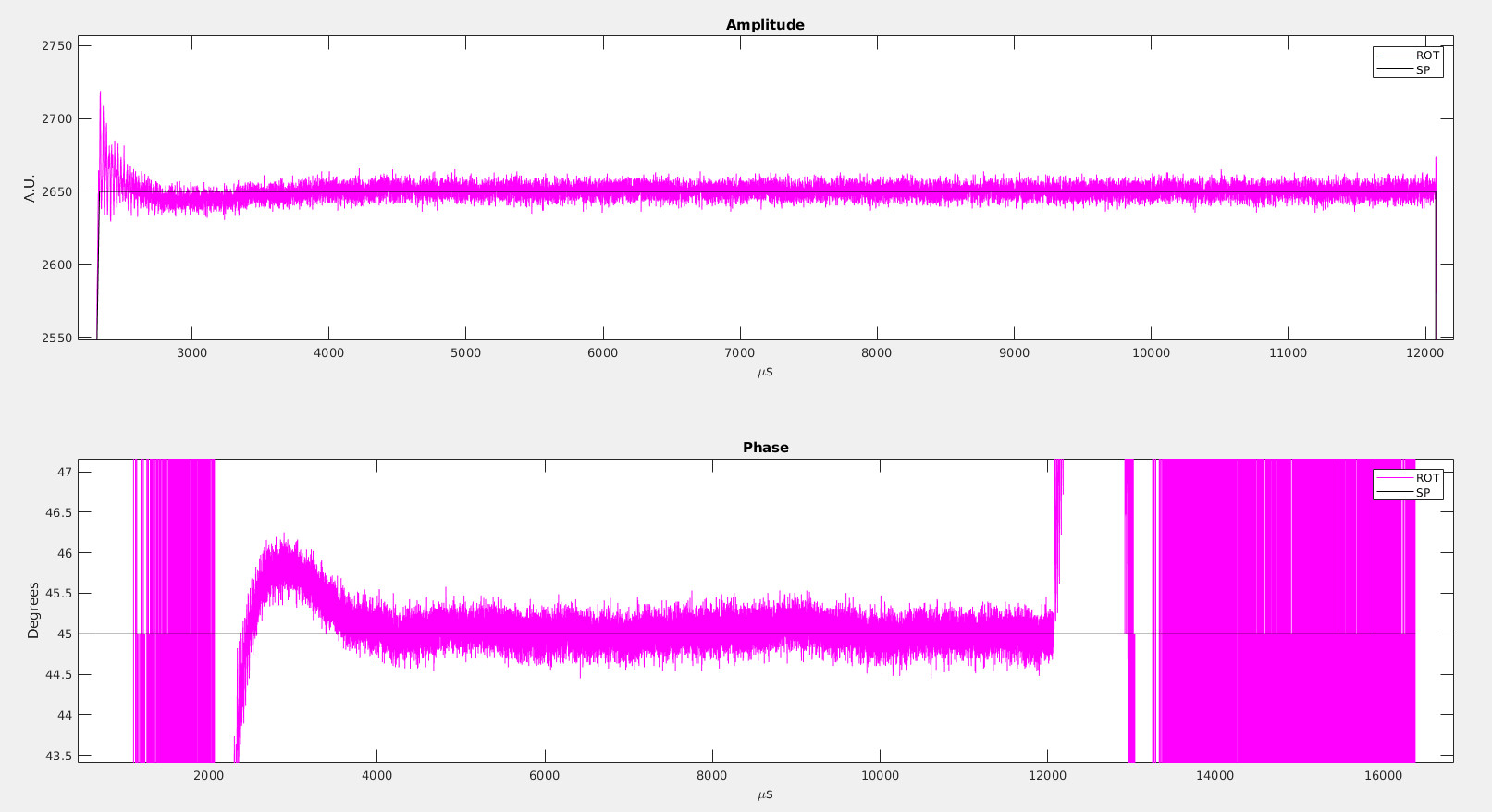}
\caption{Controller input (magnified)}
\label{fig:input_zoom}
\end{figure}

Finally, using presented test setup, feedback loop on the copper cavity has been closed.
Images below (Figs.~\ref{fig:output},\ref{fig:input}, \ref{fig:input_zoom}) shows the single pulse of the input and output of the controller.

Figure \ref{fig:output} shows amplitude and phase the output signal from the controller on top of the
feed-forward value (the ideal drive signal). The difference between feed-forward and output signal
is caused by working closed loop feedback.

Figure \ref{fig:input} shows similar image like Fig. \ref{fig:output}, but it shows the
amplitude and phase of the controller input signal on top of the set-point value (expected cavity field).

To show better the input signal on top of the set-point Figure \ref{fig:input_zoom} shows magnified amplitude
and phase regions of the RF pulse.

%
\ifboolexpr{bool{jacowbiblatex}}%
	{\printbibliography}%
	{%
	
	
} 
%
%


\end{document}